\documentclass[12pt]{article}

\usepackage{amsmath}
\usepackage{indentfirst}
\usepackage{amsfonts,amssymb}
\usepackage{mathrsfs}
\usepackage{amsthm}

\newtheorem{Definition}{Definition}[section]
\newtheorem{Theorem}{Theorem}[section]
\newtheorem{Lemma}{Lemma}[section]

\numberwithin{equation}{section}
\newtheorem{Remark}{Remark}[section]

\begin{document}
\title{{ Quasiconvex risk measures with markets volatility}
  \footnotetext{E-mail addresses: fsun.sci@outlook.com  (sunfei@whu.edu.cn)(F.Sun);
  	yjhu.math@whu.edu.cn (Y. Hu)}
  }
\author{Fei Sun, Yijun Hu}

\date{}
\maketitle
\noindent \textbf{Abstract} \indent  Since the financial markets are full of increasing volatility. In this paper, we will study the  quasiconvex risk measures defined on a special space $L^{p(\cdot)}$ where the variable exponent $p(\cdot)$ is no longer a given real number like the space $L^{p}$, but a random variable, which reflects the possible volatility of the financial markets. The dual representation for this quasiconvex risk measures will also provided.\\


\bigskip \baselineskip15pt

\section{\textbf{Introduction}}

The research of risk has become a hot topic in  the financial markets, which makes the risk management attract a great deal of attention. The measurement of finanacial risk involves two problems: the quantification of the risk in financial markets, and the allocation of that risk to individual institutions. This led to a focus on the research of risk measures.

In their seminal paper, Artzner et al. (1997, 1999) firstly introduced the class of coherent risk measures. Recent years, the quasiconvex risk measures have attracted many attentions. Cerreia-Vioglio et al.(2011) claimed that when a decision problem under uncertainty is regarded as a game against nature, the quasiconvex function can be interpreted as nature's cost function. Drapeau et al.(2011) studied the law-invariant quasiconvex risk measures. For more studies of quasiconvex risk measures, see Cerreia-Vioglio et al.(2011b), Frittelli and Maggis (2011), Drapeau and Kupper (2013), Drapeau et al.(2015), Mastrogiacomo and Gianin (2015), De-Jian and Long (2015) and references therein.

The main focus of this paper is to study a new class of quasiconvex risk measures, which was defined on a special space of financial positions, the variable exponent Bochner-Lebesgue space. Dual representation of this class of quasiconvex risk measures is provided .

The rest of the paper is organized as follows. In Section 2, we will briefly review the definition and the main properties of variable exponent Bochner-Lebesgue spaces. Section 3 is devoted to the definition of quasiconvex risk measures on the variable exponent Bochner-Lebesgue spaces.
Finally, in Section 4, we will study the dual representation of quasiconvex risk measures .\\

\section{\textbf{Preliminaries}}

In this section, we will recall the definition and the main properties of $L^{p(\cdot)}$.\\

See Cheng and Xu (2013).

\section{\textbf{Quasiconvex risk measures on $\mathbf{L^{p(\cdot)}}$ }}

 As pointed out by Cerreia-Vioglio et al.(2011), once the translation invariance is replaced with the economically sounder assumption of cash sub-additivity, the sounder mathematical translation of the `diversification' should be the so-called quasiconvexity. Especially, the cash sub-additive risk measures have been studied by Sun and Hu (2018) from the perspective of set-valued case. Further, as a special case of set-valued cash sub-additive risk measures, Sun et al.(2018) studied the set-valued loss-based risk measures.  

 In this section, the theory of quasiconvex risk measures will be extended to the markets where the volatility can not be ignored.  The main target of this section is to study the properties of quasiconvex risk measures defined on variable exponent Bochner-Lebesgue spaces.

\begin{Remark}\textnormal
{
By the definition of $L^{p(\cdot)}$, each $f\in L^{p(\cdot)}$ is a $E$-valued measurable function and $E$ is partially ordered by $K$. Hence, in the absence of ambiguity, we also regard that the $L^{p(\cdot)}$ is also partially ordered by $K$. Now, the definition of quasiconvex risk measures on $L^{p(\cdot)}$ will be introduced by axiomatic approach.
}
\end{Remark}

\begin{Definition}\label{D1}\textnormal
{
Let $E$ be a Bananch space ordered by the partial ordering relation induced by a cone $K$ and $L^{p(\cdot)}$ is a variable exponent Bochner-Lebesgue space. A mapping $\varrho:L^{p(\cdot)}\rightarrow [-\infty,+\infty]$ called quasiconvex risk measure if it satisfies
\begin{description}
  \item[A1] Monotonicity: for any $f_{1},f_{2}\in L^{p(\cdot)}$, $f_{1}\leq_{K}f_{2}\Rightarrow \varrho(f_{1})\leq \varrho(f_{2})$;
  \item[A2] Quasiconvexity: for any $f_{1},f_{2}\in L^{p(\cdot)}$ and $\lambda\in[0,1]$, $\varrho(\lambda f_{1}+(1-\lambda)f_{2})\leq \max\{\varrho(f_{1}), \varrho(f_{2})\}$.
 \end{description}
}
\end{Definition}

\begin{Remark}\textnormal
{
Note that, the quasiconvex risk measures need not satisfy the property of translation invariance, which is a key axiom for convex risk measures. Which makes the quasiconvex risk measures to be a special class of risk measures. On the other hand, the quasiconvexity also make the quasiconvex risk measures distinguish from the convex risk measures.
}
\end{Remark}

\indent In order to study the dual representation of quasiconvex risk measures, we need to introduce the concept of risk functions.

\begin{Definition}\label{D2}\textnormal
{
Let $\mathfrak{R}\Big(L^{p(\cdot)}\times \big(L^{p(\cdot)}\big)^{\ast}\Big)$ denotes the class of risk functions $R:L^{p(\cdot)}\times (L^{p(\cdot)})^{\ast}\rightarrow [-\infty,+\infty]$ that satisfy the following requirements:
\begin{description}
  \item[B1] Monotonicity: for any $f_{1},f_{2}\in L^{p(\cdot)}$ and $g\in \big(L^{p(\cdot)}\big)^{\ast} $, $f_{1}\leq_{K}f_{2}\Rightarrow R(f_{1}, g)\leq R(f_{2}, g)$;
  \item[B2] Quasiconvexity: for any $f_{1},f_{2}\in L^{p(\cdot)}$, $g\in \big(L^{p(\cdot)}\big)^{\ast} $ and $\lambda\in(0,1)$, $R(\lambda f_{1}+(1-\lambda)f_{2}, g)\leq \max\{R(f_{1}, g), R(f_{2}, g)\}$;
  \item[B3] Lower semicontinuity: $R$ is lower semicontinuous in the first component.
 \end{description}
}
\end{Definition}

Now, the acceptance sets of quasiconvex risk measures should be defined.

\begin{Definition}\label{D3}\textnormal
{
Given a quasiconvex risk measure $\varrho$, the acceptance set at level $\nu\in \mathbb{R}$ is denoted by $\mathcal{A}_{\nu}$ as follows
\begin{equation}\label{e1}
\mathcal{A}_{\nu}:=\{f\in L^{p(\cdot)}: \varrho(f)\leq \nu \}.
\end{equation}
}
\end{Definition}

\begin{Remark}\label{R1}\textnormal
{
Given a quasiconvex risk measure $\varrho$, it is easy to check that $\mathcal{A}_{\nu}$ is a closed convex set and have the monotonicity, i.e. $\nu_{1}\leq\nu_{2}$ implies $\mathcal{A}_{\nu_{1}}\subseteq\mathcal{A}_{\nu_{2}}$. In fact, by $\mathbf{A2}$, for any $f_{1},f_{2}\in \mathcal{A}_{\nu}$ and $\lambda\in[0,1]$,
\begin{displaymath}
\varrho(\lambda f_{1}+(1-\lambda)f_{2})\leq \max\{\varrho(f_{1}), \varrho(f_{2})\}.
\end{displaymath}
Since $f_{1},f_{2}\in \mathcal{A}_{\nu}$, we have $\varrho(f_{1})\leq \nu$ and $\varrho(f_{2})\leq \nu$, which implies
\begin{displaymath}
\max\{\varrho(f_{1}), \varrho(f_{2})\}\leq \nu.
\end{displaymath}
 Hence,
\begin{displaymath}
\varrho(\lambda f_{1}+(1-\lambda)f_{2})\leq \nu.
\end{displaymath}
By (\ref{e1}), we know that $\lambda f_{1}+(1-\lambda)f_{2}\in \mathcal{A}_{\nu}$, which means $\mathcal{A}_{\nu}$ is a convex set. It is also easy to show that $\mathcal{A}_{\nu}$ is a closed set and have the monotonicity.\qed
}
\end{Remark}

\begin{Lemma}\label{L1}\textnormal
{
Let $\mathcal{A}_{\nu}$ defined as Definition~\ref{D3}. Then, we have
\begin{equation}\label{e2}
f\in \mathcal{A}_{\nu} \qquad\textrm{if and only if}\qquad \langle g,f\rangle \leq \sup_{X\in\mathcal{A}_{\nu}}\langle g,X\rangle
\end{equation}
for all $g\in Q_{p(\cdot)}$ where
\begin{displaymath}
Q_{p(\cdot)}:=\Big\{g\in \big(L^{p(\cdot)}\big)^{\ast}: \frac{dg}{d\mu}\in L^{p'(\cdot)}(K_{0})\Big\}.
\end{displaymath}
}
\end{Lemma}

\noindent \textbf{Proof.}
We first show the `only if' part. If $\mathcal{A}_{\nu}=\emptyset$, the implication is obvious. If $\mathcal{A}_{\nu}\neq\emptyset$, the following implication
\begin{displaymath}
f\in \mathcal{A}_{\nu} \quad\textrm{implies}\quad \langle g,f\rangle \leq \sup_{X\in\mathcal{A}_{\nu}}\langle g,X\rangle\quad \textrm{for all } g\in Q_{p(\cdot)}
\end{displaymath}
is also straightforward.
Next, we show the `if' part. From Remark~\ref{R1}, $\mathcal{A}_{\nu}$ is a closed convex set. Thus, by Hahn-Banach theorem, for any $f\in L^{p(\cdot)}\setminus\mathcal{A}_{\nu}$, there exits a $\widehat{g}\in \big(L^{p(\cdot)}\big)^{\ast}$, such that
\begin{displaymath}
\langle \widehat{g},f\rangle > \sup_{X\in\mathcal{A}_{\nu}}\langle\widehat{g},X\rangle.
\end{displaymath}
Now, we only need to show $\widehat{g}\in Q_{p(\cdot)}$. In fact, by Remark~\ref{R24}, we have
\begin{displaymath}
\langle\widehat{g}, X\rangle=\int_{\Omega}\langle \widehat{h}, X\rangle d\mu
\end{displaymath}
where $\widehat{h}=d\widehat{g}/ d\mu\in L^{p^{'}(\cdot)}(\Omega, E^{\ast})$. Then, with the monotonicity of $\varrho$, it is easy to check $\mathcal{A}_{\nu}=\mathcal{A}_{\nu}-K$. Hence
\begin{eqnarray*}
 \langle \widehat{g},f\rangle
 > \langle \widehat{g},X-k\rangle
 &=& \int_{\Omega}\langle \widehat{h}, X-k\rangle d\mu\\
 &=& \int_{\Omega}\langle \widehat{h}, X\rangle d\mu-\int_{\Omega}\langle \widehat{h}, k\rangle d\mu\\
 &=& \langle \widehat{g},X\rangle-\int_{\Omega}\langle \widehat{h}, k\rangle d\mu
 \end{eqnarray*}
for all $k\in K$ and $X\in \mathcal{A}_{\nu}$. Thus, $\int_{\Omega}\langle \widehat{h}, k\rangle d\mu\geq0$ for all $k\in K$, which implies
$\widehat{h}\in L^{p^{'}(\cdot)}(\Omega, K_{0})$. By the definition of $Q_{p(\cdot)}$, we have $\widehat{g}\in Q_{p(\cdot)}$.\qed\\

\section{\textbf{Dual representation }}

In this section, we will study the dual representation of quasiconvex risk measures defined on variable exponent Bochner-Lebesgue spaces, which is also the main result of this paper.

\begin{Theorem}\label{T1}\textnormal
{
A mapping $\varrho:L^{p(\cdot)}\rightarrow [-\infty,+\infty]$ is a lower semicontinuous quasiconvex risk measure if and only if for any $f\in L^{p(\cdot)}$,
\begin{equation}\label{e3}
\varrho(f)=\sup_{g\in Q_{p(\cdot)}}R(f,g)
\end{equation}
where $R\in \mathfrak{R}\big(L^{p(\cdot)}\times Q_{p(\cdot)}\big)$ is expressed as
\begin{equation}\label{e5}
R(f,g):=\inf_{\nu\in \mathbb{R}}\Big\{\nu: \langle g,f\rangle \leq \sup_{X\in\mathcal{A}_{\nu}}\langle g,X\rangle\Big\}
\end{equation}
and
\begin{equation}
Q_{p(\cdot)}:=\Big\{g\in \big(L^{p(\cdot)}\big)^{\ast}: \frac{dg}{d\mu}\in L^{p'(\cdot)}(K_{0})\Big\}.
\end{equation}
}
\end{Theorem}

\noindent \textbf{Proof.}
We first show the `only if' part. Suppose $\varrho$ is a lower semicontinuous quasiconvex risk measure, we claim that $\varrho$ can be expressed as
\begin{equation}\label{e6}
\varrho(f)=\inf\big\{\nu\in \mathbb{R}: f\in \mathcal{A}_{\nu}\big\},\qquad f\in L^{p(\cdot)}.
\end{equation}
In fact, define $\varrho_{\mathcal{A}}(f):=\inf\{\nu\in \mathbb{R}: f\in \mathcal{A}_{\nu}\}$, it is easy to check that $\varrho_{\mathcal{A}}$ is a lower semicontinuous quasiconvex risk measure. Thus, we only need to show $\varrho_{\mathcal{A}}(f)=\varrho(f)$ for any $f\in L^{p(\cdot)}$. If $f\in L^{p(\cdot)}$ is such that $\varrho(f)=+\infty$, we have $\varrho_{\mathcal{A}}(f)=\varrho(f)=+\infty$. The same argumentation holds for those $f\in L^{p(\cdot)}$ satisfying $\varrho(f)=-\infty$. If $\varrho(f)\in \mathbb{R}$, we have $f\in \mathcal{A}_{\varrho(f)}$, which implies $\varrho_{\mathcal{A}}(f)\leq\varrho(f)$. On the other hand, we have $f\notin \mathcal{A}_{r}$ for any $r< \varrho(f)$. Thus, $r<\varrho_{\mathcal{A}}(f)$, which implies $\varrho(f)\leq\varrho_{\mathcal{A}}(f)$. Hence, for any $f\in L^{p(\cdot)}$
\begin{displaymath}
\varrho_{\mathcal{A}}(f)=\varrho(f)=\inf\{\nu\in \mathbb{R}: f\in \mathcal{A}_{\nu}\}.
\end{displaymath}
By Lemma~\ref{L1}, we have
\begin{equation}\label{e7}
f\in \mathcal{A}_{\nu} \qquad\textrm{if and only if}\qquad \langle g,f\rangle \leq \sup_{X\in\mathcal{A}_{\nu}}\langle g,X\rangle
\end{equation}
for all $g\in Q_{p(\cdot)}$. Then, from (\ref{e6}) and (\ref{e7}), we have
\begin{equation}\label{e8}
\varrho(f)=\inf\big\{\nu\in \mathbb{R}: \langle g,f\rangle \leq \sup_{X\in\mathcal{A}_{\nu}}\langle g,X\rangle\quad \textrm{for all } g\in Q_{p(\cdot)}\big\}.
\end{equation}
Our goal is to show that
\begin{equation}\label{e9}
\varrho(f)=\sup_{g\in Q_{p(\cdot)}}\inf_{\nu\in \mathbb{R}}\big\{\nu: \langle g,f\rangle \leq \sup_{X\in\mathcal{A}_{\nu}}\langle g,X\rangle\big\}=\sup_{g\in Q_{p(\cdot)}}R(f,g).
\end{equation}
To this end, by (\ref{e8}), we know that
\begin{displaymath}
\varrho(f)\geq\sup_{g\in Q_{p(\cdot)}}\inf_{\nu\in \mathbb{R}}\big\{\nu: \langle g,f\rangle \leq \sup_{X\in\mathcal{A}_{\nu}}\langle g,X\rangle\big\}.
\end{displaymath}
Next, we will show the reverse inequality. Suppose $\varrho(f)>-\infty$, otherwise (\ref{e9}) is trivial. Now, we fix $m<\varrho(f)$ and define
\begin{displaymath}
\mathcal{B}:=\{X\in L^{p(\cdot)}: \varrho(X)\leq m\}.
\end{displaymath}
By Remark~\ref{R1}, we know that $\mathcal{B}$ is a closed convex set and have the monotonicity. Since $f\notin \mathcal{B}$, by the Hahn-Banach theorem, there exits a $\widehat{g}\in \big(L^{p(\cdot)}\big)^{\ast}$, such that
\begin{equation}\label{e10}
\langle \widehat{g},f\rangle > \sup_{X\in\mathcal{B}}\langle\widehat{g},X\rangle.
\end{equation}
We claim that $\widehat{g}\in Q_{p(\cdot)}$. In fact, by Remark~\ref{R24}, we have
\begin{displaymath}
\langle\widehat{g}, X\rangle=\int_{\Omega}\langle \widehat{h}, X\rangle d\mu
\end{displaymath}
where $\widehat{h}=d\widehat{g}/ d\mu\in L^{p^{'}(\cdot)}(\Omega, E^{\ast})$. Then, with the monotonicity of $\varrho$, it is easy to check $\mathcal{B}=\mathcal{B}-K$. Hence, by (\ref{e10})
\begin{eqnarray*}
 \langle \widehat{g},f\rangle
 > \langle \widehat{g},X-k\rangle
 &=& \int_{\Omega}\langle \widehat{h}, X-k\rangle d\mu\\
 &=& \int_{\Omega}\langle \widehat{h}, X\rangle d\mu-\int_{\Omega}\langle \widehat{h}, k\rangle d\mu\\
 &=& \langle \widehat{g},X\rangle-\int_{\Omega}\langle \widehat{h}, k\rangle d\mu
 \end{eqnarray*}
for all $k\in K$ and $X\in \mathcal{B}$. Thus, $\int_{\Omega}\langle \widehat{h}, k\rangle d\mu\geq0$ for all $k\in K$, which implies
$\widehat{h}\in L^{p^{'}(\cdot)}(\Omega, K_{0})$. By the definition of $Q_{p(\cdot)}$, we have $\widehat{g}\in Q_{p(\cdot)}$.\\
For all $\nu\leq m$,  we have $\mathcal{A}_{\nu}\subseteq\mathcal{B}$. Then
\begin{equation}\label{e11}
\sup_{X\in \mathcal{A}_{\nu}}\int_{\Omega}\langle \widehat{h}, X\rangle d\mu\leq \sup_{X\in \mathcal{B}}\int_{\Omega}\langle \widehat{h}, X\rangle d\mu.
\end{equation}
Thus, by (\ref{e10}) and (\ref{e11})
\begin{equation}\label{e12}
\langle \widehat{g},f\rangle-\sup_{X\in \mathcal{A}_{\nu}}\langle \widehat{g},X\rangle\geq \langle \widehat{g},f\rangle-\sup_{X\in \mathcal{B}}\langle \widehat{g},X\rangle>0.
\end{equation}
Since for each $\nu\leq m$, we can imply (\ref{e12}) and by the fact that $\nu\mapsto \sup_{X\in \mathcal{A}_{\nu}}\langle g,X\rangle$ is nondecreasing, we have
\begin{equation}\label{e13}
m\leq\sup_{g\in Q_{p(\cdot)}}\inf_{\nu\in \mathbb{R}}\big\{\nu: \langle g,f\rangle \leq \sup_{X\in\mathcal{A}_{\nu}}\langle g,X\rangle\big\}.
\end{equation}
This relation holds for each $m<\varrho(f)$. Hence
\begin{displaymath}
\varrho(f)\leq\sup_{g\in Q_{p(\cdot)}}\inf_{\nu\in \mathbb{R}}\big\{\nu: \langle g,f\rangle \leq \sup_{X\in\mathcal{A}_{\nu}}\langle g,X\rangle\big\}.
\end{displaymath}
Then,
\begin{equation}\label{e14}
\varrho(f)=\sup_{g\in Q_{p(\cdot)}}R(f,g).
\end{equation}
Now, we only need to show $R\in \mathfrak{R}\big(L^{p(\cdot)}\times Q_{p(\cdot)}\big)$. First, by the monotonicity and lower semicontinuity of $\varrho$ with (\ref{e14}), it is easy to check that $R$ satisfies $\mathbf{B1}$ and $\mathbf{B3}$. Next, we will show that $R$ satisfies $\mathbf{B2}$. By Remark~\ref{R24}, we have
\begin{displaymath}
\langle g, X\rangle=\int_{\Omega}\langle h, X\rangle d\mu
\end{displaymath}
where $h=d g/ d\mu\in L^{p^{'}(\cdot)}(\Omega, E^{\ast})$. For any $f_{1},f_{2}\in L^{p(\cdot)}$, $\alpha\in(0,1)$ and $g\in Q_{p(\cdot)}$,
\begin{eqnarray*}
 R(\alpha f_{1}+(1-\alpha)f_{2}, g)
 &=& \inf_{\nu\in \mathbb{R}}\Big\{\nu: \langle g,\alpha f_{1}+(1-\alpha)f_{2}\rangle \leq \sup_{X\in\mathcal{A}_{\nu}}\langle g,X\rangle\Big\}\\
 &=& \inf_{\nu\in \mathbb{R}}\Big\{\nu: \int_{\Omega}\langle h,\alpha f_{1}+(1-\alpha)f_{2}\rangle  d\mu \leq \sup_{X\in\mathcal{A}_{\nu}}\langle g,X\rangle\Big\}\\
 &=& \inf_{\nu\in \mathbb{R}}\Big\{\nu: \alpha\int_{\Omega}\langle h, f_{1}\rangle d\mu +(1-\alpha)\int_{\Omega}\langle h, f_{2}\rangle d\mu \leq \sup_{X\in\mathcal{A}_{\nu}}\langle g,X\rangle\Big\}\\
 &=& \inf_{\nu\in \mathbb{R}}\Big\{\nu: \alpha\langle g,f_{1}\rangle +(1-\alpha)\langle g,f_{2}\rangle  \leq \sup_{X\in\mathcal{A}_{\nu}}\langle g,X\rangle\Big\}.\\
 \end{eqnarray*}
Without loss of generality, let $\langle g,f_{1}\rangle\geq \langle g,f_{2}\rangle$. Then
\begin{eqnarray*}
R(\alpha f_{1}+(1-\alpha)f_{2}, g)&\leq&\inf_{\nu\in \mathbb{R}}\Big\{\nu: \langle g,f_{1}\rangle\leq \sup_{X\in\mathcal{A}_{\nu}}\langle g,X\rangle\Big\}\\
&=&R(f_{1},g)\\
&\leq&\max\{R(f_{1},g),R(f_{2},g)\},
\end{eqnarray*}
which means $R$ satisfies $\mathbf{B2}$. Therefore, $R\in \mathfrak{R}\big(L^{p(\cdot)}\times Q_{p(\cdot)}\big)$.\\
\indent Now, we will show the `if' part. Suppose that $\varrho(f)=\sup_{g\in Q_{p(\cdot)}}R(f,g)$ for a risk function  $R\in \mathfrak{R}\big(L^{p(\cdot)}\times Q_{p(\cdot)}\big)$ where
$R(f,g)=\inf_{\nu\in \mathbb{R}}\big\{\nu: \langle g,f\rangle \leq \sup_{X\in\mathcal{A}_{\nu}}\langle g,X\rangle\big\}$. The properties of monotonicity and lower semicontinuity of $\varrho$ are the direct consequences of $\mathbf{B1}$ and $\mathbf{B3}$. Now, we only need to show that $\varrho$ satisfies $\mathbf{A2}$. Since $R$ satisfies $\mathbf{B2}$, for any $\lambda\in (0,1)$ and $f_{1},f_{2}\in L^{p(\cdot)}$,
\begin{equation}\label{e15}
R(\lambda f_{1}+(1-\lambda)f_{2}, g)\leq \max\{R(f_{1}, g), R(f_{2}, g)\}.
\end{equation}
Thus, it follows that
\begin{eqnarray*}
\varrho(\lambda f_{1}+(1-\lambda)f_{2})&=&\sup_{g\in Q_{p(\cdot)}}R(\lambda f_{1}+(1-\lambda)f_{2}, g)\\
&\leq&\sup_{g\in Q_{p(\cdot)}}\max\{R(f_{1}, g), R(f_{2}, g)\}\\
&\leq&\max\{\varrho(f_{1}), \varrho(f_{2})\}.
\end{eqnarray*}
Hence, $\varrho$ is a lower semicontinuous quasiconvex risk measure.\qed\\

%\end{CJK*}
\end{document}